# Observation of the possible chiral edge mode in Bi$_{1-x}$Sb$_x$

M. Sasaki[1], A. Ohnishi[1], Nabyendu Das[2], K.-S. Kim[2,*], and Heon-Jung Kim[3,4,†]

[1] Department of Physics, Faculty of Science, Yamagata University, Kojirakawa, Yamagata 990-8560, Japan

[2] Department of Physics, Pohang University of Science and Technology, Pohang, Gyeongbuk 37673, Republic of Korea

[3] Department of Materials-Energy Science and Engineering, College of Engineering, Daegu University, Gyeongbuk 38453, Republic of Korea.

[4] Department of Physics, College of Natural and Life Science, Daegu University, Gyeongbuk 38453, Republic of Korea

E-mail: †hjkim76@daegu.ac.kr, *tkfkd@postech.ac.kr

**Abstract**

After the classification of topological states of matter has been clarified for non-interacting electron systems, the theoretical connection between gapless boundary modes and nontrivial bulk topological structures, and their evolutions as a function of dimensions are now well understood. However, such dimensional hierarchy has not been well established experimentally although some indirect evidences were reported, for example, such as the half-quantized Hall conductance via quantum Hall effect and extrapolation in the quantum-oscillation measurement. In this paper, we report the appearance of the possible chiral edge mode from the surface state of topological insulators under magnetic fields, confirming the dimensional hierarchy in three dimensional topological insulators. Applying laser pulses to the surface state of Bi$_{1-x}$Sb$_x$, we find that the sign of voltage relaxation in one edge becomes opposite to that in the other edge only when magnetic fields are applied to the topological insulating phase. We show that this sign difference originates from the chirality of edge states, based on coupled time-dependent Poisson and Boltzmann equations.

---

[*] Co-corresponding author
[†] Co-corresponding author

Chiral edge states are boundary states emerging in two-dimensional (2D) electron systems and reflect its bulk topology. These exotic states have been observed in quantum Hall systems [1-4], graphene [5,6], HgTe [7,8], surface of three-dimensional (3D) topological insulators (TI) [9,10,11], and quantum anomalous Hall systems [12,13]. In fact, prerequisite for observing the edge states has been thought to be to apply high perpendicular magnetic field $B$ so that well-separated Landau levels are established. When the Fermi energy $E_F$ is placed in between the bulk Landau gap, as a consequence of band bending, there are states crossing it at the edges. In contrast to usual quantum Hall systems, 2D Dirac electron systems exhibits half-integer quantum Hall effect, as first observed in graphene [5,6] and later in HeTe [7,8]. The Landau levels of 2D Dirac fermions have $sign(N)v_F\left(2eB\hbar|N|\right)^{1/2}$, where $sign$ is the sign function, $N$ is the Landau index, $v_F$ is the Fermi velocity, $e$ is the electron charge, $\hbar$ is Planck constant divided by $2\pi$. This is in contrast to usual Landau levels $\hbar\omega_c(N+1/2)$ and leads to half-quantized Hall conductance $\sigma_{xy} = g(N+1/2)e^2/\hbar$, where $g$ is the number of degenerate species of Dirac fermions. The half-quantized Hall conductance is expected to occur in 3D TI. In this case, interestingly, a hierarchy structure is observable, starting from the bulk 3D TI to one-dimensional (1D) edge state through topological 2D surface state. Topological 2D surface state is a boundary state of the bulk 3D TI and 1D edge state occurs as a boundary state of the topological 2D surface state when gap opens at the Dirac node and $E_F$ of the 2D surface state is placed at the node.

In general, in the topological state of matter [14-18], there exist structures called bulk-edge correspondence between gapless boundary modes and nontrivial bulk topological structures, and their evolutions as a function of dimensions. This dimensional hierarchy shows how gapless boundary states evolve as dimensional reduction is performed. Theoretically, the existence of gapless surface states in three-dimensional topological insulators is attributed to the topological $\vec{E}\cdot\vec{B}$ term with an angle coefficient $\theta = \pi$ [19-23]. For the two-dimensional surface state, this term develops into the Chern-Simons term when time reversal symmetry is broken by the applied magnetic fields [19-23]. The half-quantized Hall conductance is a fingerprint of this dimensional hierarchy that starts from topological insulators in 3D [24]. Unfortunately, the half-quantized Hall conductance could be seen only by extrapolation in the quantum-oscillation measurement [25-27]. Unambiguous evidence is to observe the single chiral edge mode more directly, responsible for the half-quantized Hall conductance. This was achieved only very recently in 3D TIs [9,10,11].

In this paper, we try to understand the bulk-boundary correspondence in $Bi_{1-x}Sb_x$ by performing specially designed transient experiments, which leads to the surface Dirac electrons with a gap. According to the dimensional hierarchy, the single chiral edge mode should appear at least in the region of weak magnetic fields, which results from the single Dirac cone topologically protected by the $Z_2$ index for the surface mode. It is a main point of this study to observe such a chiral edge state by applying laser pulses to the surface state and scrutinizing the relaxation of voltage drop in each one-dimensional edge. We find that the sign of voltage drop in one side, saying $V_{4\to 3}$, becomes opposite to that in the other side, saying $V_{6\to 5}$, i.e., $\text{sgn}(V_{4\to 3}) = -\text{sgn}(V_{6\to 5})$ only when magnetic fields are applied to topological insulators [see Fig. 1(b)]. On the other hand, $\text{sgn}(V_{4\to 3}) = \text{sgn}(V_{6\to 5})$ is gained when (no) magnetic fields are applied to a topologically (non-) trivial state. We suggest that this sign difference originates from the existence of the chiral edge mode in a time-reversal symmetry-broken topological insulating phase by demonstrating that both the sign and time dependence of the voltage drop are explained even quantitatively in the framework of coupled time-dependent Poisson and Boltzmann equations based on the physical picture of the chiral edge mode.

To detect such a chiral edge mode, we devise measurements at $T$ = 4.2 K of transient thermoelectric voltages induced by the Nd-YAG pulsed laser with wavelength of 1064 nm [28] in a magnetic field. The energy density on the sample is ~ 15 mJ/cm$^2$. We call this experiment M1 hereafter, where the sample is covered half to block the laser light [Fig. 1(b)]. The metal mask was manually made by Al foils covered with Kapton tape. The Kapton tape was used for electrical insulation. The experimental setup and dimension of the samples in the M1 experiment are illustrated schematically in

Fig. 1(b). Typical thickness of the sample is around 0.5 mm. The exposure area and mask size are 3 mm$^2$ and 6 mm$^2$, respectively. The laser pulse instantaneously causes the distribution of electrons to deviate from equilibrium. In a metal, non-equilibrium temperatures of electrons are produced and electrons are equilibrated with phonons within few picoseconds through electron-phonon coupling [29,30]. In the case of a semiconductor, the laser pulse generates electron-hole pairs also within few picoseconds [31]. These processes eventually increase the temperature of an entire system. The increase of temperature is expected to be about 10 K, evaluated from the delivered energy from the laser pulse and the specific heat of the BiSb alloy. Thus, the laser irradiation causes the temperature of the system higher in the exposed part of the sample at the very initial stage. In our experiments, after the laser irradiation, we investigate the equilibrating process by measuring the temporal evolution of the voltage differences $V_{4\to3}$ and $V_{6\to5}$ on each side of the sample with a high-speed oscilloscope having resolution of approximately one nanosecond. Here 3, 4, 5, and 6 are contact numbers and the contact 6 and 4 are grounded. In the M1 set-up, the contact 3 and 5 are located in the region covered by the mask, and the contact 4 and 6 are in the uncovered part as shown in Fig. 1(b).

In this experiment, we used $Bi_{1-x}Sb_x$ single crystals with $x$ = 0.0 %, 3.0 %, 11.6 %, and 21.5 %. We selected these samples taking account for the phase diagram of the BiSb alloy and their gap values [32]. It is well-known that the topological phase transition from a band insulator to a topological insulator occurs across $x \sim$ 3 - 4 % in $Bi_{1-x}Sb_x$ [32,33]. In this special composition, this system evolves into a Weyl metal state under the magnetic field, and indeed negative longitudinal magnetoresistance [34] and violation of Ohm's law [35] originated in chiral anomaly of the Weyl metallic state were observed. However, the system in this composition is still metallic due to contributions of the hole band around the T point in the reciprocal space. A full insulating state appears above $x \sim$ 7 % when the band at T sinks below the Fermi level. In this respect one may regard Bi and $Bi_{1-x}Sb_x$ with $x$ = 3.0% as band "insulators" while $Bi_{1-x}Sb_x$ with $x$ = 11.6% and 21.5% as topological insulators. The temperature dependence of resistivity informs us that our samples for x < 3.0 % are metallic with residual resistivity at $T = 4.2$ K less than 0.03 mΩcm while those with $x > 7.0$ % are weakly insulating with residual resistivity of 0.7 – 2.1 mΩcm. Fig. 2 displays $V_{4\to3}(t)$ and $V_{6\to5}(t)$ in the M1 configuration for $Bi_{1-x}Sb_x$ single crystals with $x = 0.0$ %, 3.0 %, and Fig. 3 for $x = 11.6$ %, and 21.5 %.

Several important points are clearly noticed in these figures. First, the transient voltage signals rise with a characteristic time less than 0.05 μsec. The signals reach the maximum and decay with characteristic time dependence. We will discuss the characteristic time scales in detail later. In fact, the time scales of the rise and decay are much larger than the time constant of electronic processes measured by the femto-second experiments, which is time scale of only few picoseconds [29-31]. The maximum signal is proportional to the magnetic field in the low-field region, but it is saturated or slightly decreases at higher magnetic fields around 1-4 T. The polarity of the signal is determined by the sign of the applied magnetic field and thus, the sign of the voltage signal changes when the magnetic field is reversed. The most interesting observation is the sign of $V_{4\to3}(t)$ and $V_{6\to5}(t)$ under a given magnetic field. It turns out to be the same in the band insulating region but it is opposite in the topological insulating region. In the case of Bi, Fig. 2(a) and (b) show negative signs of $V_{4\to3}(t)$ and $V_{6\to5}(t)$ for $B > 0$ and positive signs for $B < 0$. In contrast, as shown in Fig. 3, the sign of $V_{4\to3}(t)$ is opposite to that of $V_{6\to5}(t)$ in the topological insulating state. Compare Fig. 3(a) [$V_{4\to3}(t)$] with Fig. 3(b) [$V_{6\to5}(t)$] for the $x = 11.6$ % sample and Fig. 3(c) [$V_{4\to3}(t)$] with Fig. 3(d) [$V_{6\to5}(t)$] for the $x = 21.5$ % sample. As the Bi and $Bi_{1-x}Sb_x$ with $x = 3.0$ % possess quite different band structure, the peculiar band structure is not a key factor to give rise to the same sign in the band insulating region.

The observation in the band insulating region indicates that the transient current flows from the region at high temperature to that at low temperature as expected. However, the sign difference in the topological insulating region implies that the transient current circulates in either clockwise or counter-clockwise, depending on magnetic fields [Fig. 1(a)] in contrast to the case of the band insulator. We attribute the origin of this sign difference to the possible appearance of chiral edge modes on the surface of a topological insulator under magnetic fields. To further support this scenario, we devised another experiment, which was also performed at $T = 4.2$ K with a different mask configuration as shown in Fig. 1(c). This configuration, which we call the M2 or differential, cancels non-circulating currents, and thus it maximizes the circulating ones. In this setup, all voltage contacts are screened by two masks while the middle of the sample is open. The schematic diagram of experimental setup are presented, along with the sample dimensions and exposure area in Fig. 1(c). In case that there are no 1D chiral edge modes and the sample-mask-contact configuration is perfectly symmetric, the voltage 3 (5) and 4 (6) should be exactly same, giving the zero-voltage difference between 3 (5) and 4 (6). Because the real experimental setup is not perfectly symmetric, a small signal due to experimental asymmetry may appear. However, its magnitude is expected to be smaller than the corresponding signal in the M1 configuration.

Figures 4(a), 4(b), 4(c), and 4(e) display $V_{4\to3}(t)$ of $Bi_{1-x}Sb_x$ with $x = 0$ %, 3 %, 11.6 %, and 21.5 %, respectively, in the M2 configuration. Compared to M1, the background signals are observed to decrease. $V_{4\to3}(t)$ for $Bi_{1-x}Sb_x$ with $x = 0$ % and 3 % is much reduced on the whole compared to M1, whose maximum is only 1/4 of that in M1. In contrast, $V_{4\to3}(t)$ for topological insulating samples becomes more enhanced overall, compared to the case of M1. The magnitude of the peak signal is two times larger than that in M1. In addition, the magnitude of $V_{6\to3}(t)$ is quite reduced in topological insulating samples. $V_{4\to3}(t)$ and $V_{6\to3}(t)$ in the M2 configuration also support the existence of the possible circulating current in the topological insulating state under magnetic fields in stark contrast with the case of the band insulating state of $Bi_{1-x}Sb_x$. All observed signals in our experiments are expected to come from the top surface predominantly because only the top surface is directly irradiated by the laser pulse and the area of the side surface is quite small compared to the top. For comparison purpose, we performed the M1 and M2 experiments on a n-doped GaAs single crystal. This sample shows main characteristic features of a band insulator. First, signs of $V_{4\to3}(t)$ and $V_{6\to3}(t)$ are same in the M1 experiment [Fig. 5(a) and Fig. 5(b)]. Second, the signal is reduced in the M2 experiment compared to the one in the M1 experiment. For this, please compare Fig. 5(a) and Fig. 5(c).

To justify that the chiral-edge mode gives rise to the transient circulating current, we resort to the coupled time-dependent Poisson and Boltzmann equations, where the time-dependent electric potential $\phi$ or voltage relaxation is described by the Maxwell equation, where $\phi$ satisfies

$$\partial_x^2 \phi_{R(L)}(x,t) - \partial_t^2 \phi_{R(L)}(x,t) = \frac{q}{\varepsilon} \int_{-\infty}^{\infty} dv \int_{-\infty}^{\infty} d\omega \{f_{R(L)}^+(x,t;v,\omega) - f_{R(L)}^-(x,t;v,\omega)\},$$

and the relaxation dynamics of the chiral mode is described by the Boltzmann equation,

$$\partial_t f_{R(L)}^\pm(x,t;v,\omega) + v_{R(L)} \partial_x f_{R(L)}^\pm(x,t;v,\omega) \pm \frac{q}{m}[E_s(x,t) - \partial_x \phi_{R(L)}(x,t)] \partial_v f_{R(L)}^\pm(x,t;v,\omega)$$
$$\pm q[\partial_t \phi_s(x,t) + \partial_t \phi_{R(L)}(x,t)] \partial_\omega f_{R(L)}^\pm(x,t;v,\omega) = 0$$

where $f_{R(L)}^\pm(x,t;v,\omega)$ is a distribution function. Here, $R(L)$ in the subscript of both the electric potential and distribution function represents the right (left) chiral edge mode, and $+(-)$ in the superscript of the distribution function denotes the distribution of positive (negative) charges. $q$ and $\varepsilon$ in the Poisson equation are an electric charge and a dielectric constant of the surface state, respectively. $m$ in the Boltzmann equation is the mass of the chiral electron. There are two essential points in this Boltzmann equation. One is that the sign of the velocity of the right propagating mode in one side is opposite to that of the left propagating mode in the other side, i.e., $v_R = +v$ and $v_L = -v$, representing the chirality of such an emergent one-dimensional mode under external magnetic fields. The other point is that there is no dissipation term on the right hand side of the Boltzmann equation, which indicates robustness of the 1D chiral edge mode against disorder. Temporal and spatial evolution of the applied laser pulse can be approximated with $E_s(x,t) = \frac{E}{2} \text{sech}^2\left(\frac{t-t_0}{\tau}\right)\left\{1 + \tanh\left(\frac{x}{\xi}\right)\right\}$, resulting in

$\phi_s(x,t) = -\frac{E}{2}\text{sech}^2\left(\frac{t-t_0}{\tau}\right)\left\{x + \ln\cosh\left(\frac{x}{\xi}\right) + \xi\ln 2\right\}$ for an appropriate gauge choice, where $\tau$ is full width at half maximum of the laser pulse and $\xi$ is a length scale related with the laser spot size.

Solving the Boltzmann equation, we obtain $f_{R(L)}^{\pm}(x,t;v,\omega) = n\exp\left\{-\beta\left(\frac{mv^2}{2} \pm \text{sgn}(v_{R(L)})q[\phi_s(x,t) + \phi_{R(L)}(x,t)] - \text{sgn}(v_{R(L)})\omega\right)\right\}$, where $n$ is the density of surface electrons and $\beta$ is an inverse temperature. We point out that the sign difference appears in this expression due to the chirality. Inserting this distribution function into the Maxwell equation, we reach the following expression $\partial_x^2\phi_{R(L)}(x,t) - \partial_t^2\phi_{R(L)}(x,t) = -\text{sgn}(v_{R(L)})\frac{2nq}{\beta\varepsilon}C\sinh\{\beta q[\phi_s(x,t) + \phi_{R(L)}(x,t)]\}$, where $C$ is a positive constant that results from the velocity integration.

As this is a nonlinear equation, it is not easy to solve. A usual way is to linearize this equation, expanding the sinh-term up to the first order in the electric potential. Then, an asymptotic solution is given by $\phi_{R(L)}(x,t) = \Phi_{R(L)}^{(0)}(x) + \exp\left(-2\frac{t-t_0}{\tau}\right)\Phi_{R(L)}^{(1)}$ in the $t \to \infty$ limit. The first term represents an offset associated with the boundary condition, which is not important here. The second term describes the relaxation dynamics of voltage drop. A crucial point is that the relaxation time scale is purely given by the pulse width of the right hand side in the Maxwell equation. This is a crucial result for the 1D chiral fermion, directly originating from the absence of the dissipation term on the right hand side of the Boltzmann equation. It is also clear that the sign difference in the voltage drop results from the sign difference of the velocity of the chiral mode on each side.

In Fig. 6, we present the analysis of $V_{4\to 3}(t)$ for topological insulating samples in the M2 configuration. According to the analytic solution in the Maxwell-Boltzmann equation approach above, both the initial rise and the final decay in $V_{4\to 3}(t)$ may be parameterized by the same functional form as the electric field of the laser pulse, that is, the source term, whose temporal dependence is given by

$\text{sech}^2\left(\dfrac{t-t_0}{\tau}\right)$. The red lines in Fig. 6(a) and (b) represent our theoretical fitting curves to the $\text{Bi}_{1-x}\text{Sb}_x$ data normalized by the maximum value. As shown in Fig. 6(a), this functional form fits the data even quantitatively well over the entire time scale for $\text{Bi}_{1-x}\text{Sb}_x$ with $x = 11.6\,\%$, giving $\tau \sim 14$ $n$sec and $t_0 \sim 30-40$ $n$sec. On the other hand, for $\text{Bi}_{1-x}\text{Sb}_x$ with $x = 21.5\,\%$, though the initial part of the experimental data is quite well reproduced by the theoretical curve with $\tau \sim 20$ $n$sec and $t_0 \sim 40-45$ $n$sec, the data beyond the peak region deviate significantly, indicating another relaxation mechanism in operation. It is worth noting that $\tau$ of both samples are in the same magnitude with the pulse width of the laser (~ 10 $n$sec), consistent with our Maxwell-Boltzmann equation approach based on the chiral edge mode.

To determine the second relaxation time in $\text{Bi}_{1-x}\text{Sb}_x$ with $x = 21.5\,\%$, we analyse the data beyond the maximum, based on $V_{4\to 3}(t) = V_s \exp(-t/\tau_s) + V_l \exp(-t/\tau_l) + V_0$, where $\tau_s$ and $\tau_l$ are shorter and longer relaxation times, respectively. The result of this analysis is presented in fig. 6(c), giving $\tau_s \sim 20$ $n$sec. $\tau_s$ is same with $\tau$ as it should be and $\tau_l \sim 0.1-0.4$ μsec. The origin of this second relaxation time is not clear at this moment. Comparing this relaxation time with that in pure Bi samples, we suspect that the chiral edge mode in $\text{Bi}_{1-x}\text{Sb}_x$ with $x = 21.5\,\%$ is more strongly coupled to bulk electronic degrees of freedom than the case of $\text{Bi}_{1-x}\text{Sb}_x$ with $x=11.6\,\%$. We speculate that stronger couplings between the chiral edge mode and bulk degrees of freedom in $\text{Bi}_{1-x}\text{Sb}_x$ with $x = 21.5\,\%$ may originate from larger localization length into the bulk from the edge than that of $\text{Bi}_{1-x}\text{Sb}_x$ with $x = 11.6\,\%$, implying that $\text{Bi}_{1-x}\text{Sb}_x$ with $x=11.6\,\%$ has stronger topological properties than $\text{Bi}_{1-x}\text{Sb}_x$ with $x = 21.5\,\%$.

In Fig. 6(d), we analyse the scaled $V_{4\to 3}(t)$ in the M2 configuration with $\text{sech}^2\left(\dfrac{t-t_0}{\tau}\right)$ for Bi single crystals, where our fitting produces $\tau \sim 25$ $n$sec and $t_0 \sim 40$ $n$sec. It is now natural to understand $\text{sgn}(V_{4\to 3}) = \text{sgn}(V_{6\to 5})$, where chiral edge states do not form on the Bi surface under magnetic fields. However, the reason is still unclear why the relaxation time scale in Bi is more enhanced than that in $Bi_{1-x}Sb_x$ with $x = 11.6$ %. We suspect that this may originate from some types of scattering effects in surface electrons. In order to introduce scattering in surface electrons, we consider a conventional relaxation term in the right hand side of the Boltzmann equation with a scattering time $\tau_{sc}$, where it would result from scatterings between surface electrons themselves or between surface electrons and bulk charge carriers. Solving the Boltzmann equation with scattering and inserting the solution into the Maxwell equation, we reach the following time-dependent Poisson equation,

$$\nabla^2 \phi(x,t) - \partial_t^2 \phi(x,t) = -\frac{2nq^2}{\varepsilon} C \left\{ \phi_s(x,t) + \phi(x,t) + \beta \tau_{sc}^2 \frac{q^2}{2m^3} \left( E_s(x,t) - \nabla \phi(x,t) \right)^2 \right\}$$

, where the exponential Boltzmann factor is "linearized". Although it is extremely difficult to solve this nonlinear differential equation, one can expect that the relaxation time becomes enlarged because of scattering processes, i.e., $\tau_{sc} \neq 0$.

The inset of Fig. 6(d) implies three steps of relaxation processes, which correspond to three different time scales; $\tau$ estimated above, the oscillation period, and the period related with damping of this oscillation. Similar experiments on several *n*- and *p*-doped semiconductors without magnetic fields also confirmed existence of three steps of relaxation processes; the carrier generation and recombination, the diffusion of carriers, and the diffusion of thermal flux or phonons along the temperature gradient [36-38]. Even in the presence of external magnetic fields, these three processes will still be main mechanisms for the relaxation processes. The carrier mobility would be affected by the cyclotron motion under magnetic fields. We leave sincere analysis for Bi as an interesting future study. As the analysis in Fig. 6(d) is also equally well applied to $Bi_{1-x}Sb_x$ with $x = 3$ %, three relaxation processes are expected to exist in this sample. It is also noted that the peak signal at $B = 4$ T in the M1 experiment of Bi is much larger than that of other compositions. As the M1 experiment is a dynamic version of thermoelectric measurement, this peak signal is proportional to thermoelectric coefficient. This implies that Bi has larger thermoelectric coefficient than $Bi_{1-x}Sb_x$ with $x = 3$ %. Therefore, by accurately interpreting the time scales discussed above, the M1 experiment can provide important clues to understand and optimize thermoelectric performance of any thermoelectric material including Bi and BiSb alloy, particularly in the band "insulating" region.

Finally, we emphasize that the rather precise scaling behaviour from 0.4 T to 4 T implies the existence of only one single chiral-edge mode. This is consistent with the expectation that one single chiral-edge mode should exist on the surface state in the weak-field limit. On the other hands, strong magnetic fields are expected to lead to a series of Landau levels on the surface state. In fact, such Landau levels on the surface of 3D TI were directly measured in the dual-gated $BiSbTeSe_2$ devices [11]. The independent, ambipolar gating of parallel-conducting top and bottom surfaces allows for observations of twice the conductance quantum at zero field, a series of two-component half-integer Dirac quantum Hall states, and an electron-hole total filling factor zero state. It is interesting to investigate the relaxation dynamics of these 2D Dirac fermion states involving various combinations of top and bottom surface half-integer filling factor. Whether their relaxation dynamics depends on the fill factor is an important and unaddressed problem at present.

In conclusion, we devised measurements of transient thermoelectric voltages induced by the Nd-YAG pulsed laser. Laser pulses cause the opposite sign of the transient voltage on each side contact of the surface state in topological insulators when dimension is reduced by applying external magnetic fields. Resorting to the Maxwell-Boltzmann equation approach, we could verify that this sign difference originates from the chirality of edge states. Our relaxation experiment shed light on the nature of delocalized states, resulting from the nontrivial bulk topological structure. Our study confirms the dimensional hierarchy or the bulk-boundary correspondence of the AII class, starting from three dimensional topological insulators.

**Acknowledgements**

This study was supported by Basic Science Research Program through the National Research Foundation of Korea (NRF) funded by the Ministry of Science, ICT & Future Planning (2017R1A2B2002731). K.-S.K. was supported by the Ministry of Education, Science, and Technology (No. NRF-2015R1C1A1A01051629 and No. 2011-0030046) of the National Research Foundation of Korea (NRF) and by the POSTECH Basic Science Research Institute Grant (2017). K.-S.K appreciated great hospitality of Asia Pacific Center for Theoretical Physics (APCTP), where this work was initiated. We would like to express our thanks to Professor T. Iwata of Faculty of Science, Yamagata University for his encouragement.

**Figure caption**

Fig. 1 (a) The schematic picture is shown about one-dimensional edge modes, which can sustain the circulating currents in a magnetic field. The direction of the applied magnetic field selects its chirality. 1 (b) In the upper part is depicted the M1 experiment setup. The laser light induces transient voltage signals $V_{4\to3}(t)$ and $V_{6\to5}(t)$, where the contact 4 and 6 are grounded. In the lower part is shown approximate sample dimensions and configuration in the M1 experiment, along with exposure area. (c) In the upper is presented the schematic diagram of the M2 mask configuration, where the middle of the sample is exposed to the laser light while remaining parts are screened by metal masks. In the lower part is shown sample dimensions and configuration in the M2 experiment, along with exposure area.

Fig. 2 These data are the results of the M1 experiments for $Bi_{1-x}Sb_x$ with $x = 0$ % and 3 %. The experimental $V_{4\to3}(t)$ and $V_{6\to5}(t)$ of $Bi_{1-x}Sb_x$ with $x = 0$ % [$x = 3.0$ %] single crystal, which lies in the band "insulating" region, are presented in (a) [(c)] and (b) [(d)], respectively. The sign of $V_{4\to3}(t)$ and $V_{6\to5}(t)$ changes when the polarity of the magnetic field is reversed. In a given magnetic field, the signs of $V_{4\to3}(t)$ and $V_{6\to5}(t)$ are same.

Fig. 3 These are the results of the M1 experiments for $Bi_{1-x}Sb_x$ with $x = 11.6$ % and 21.5 %. The sign of $V_{4\to3}(t)$ is opposite to that of $V_{6\to5}(t)$ in a given magnetic field for $Bi_{1-x}Sb_x$ with $x = 11.6$ % [(a)-(b)] and 21.5 % [(c)-(d)], which are in the topological insulating states. This evidences existence of the circulating current.

Fig. 4 Experimental $V_{4\to3}(t)$ data of $Bi_{1-x}Sb_x$ with x = 0 % and 3 % are shown for $B > 0$ and $B < 0$ in (a) and (b), respectively. In case of Bi, the magnitude of the peak in $V_{4\to3}(t)$ (~ 50 meV) is less than one fourth of the corresponding peak value in the M1 configuration (~ 200 meV). The residual signal may occur due to the asymmetry in the mask-sample configuration. Similar behaviour can be seen in $Bi_{1-x}Sb_x$ with x = 3 %. In contrast, for $Bi_{1-x}Sb_x$ with $x$ = 11.6 % [(c)] and $x$ = 21.5 % [(e)], the magnitudes of the peak in $V_{4\to3}(t)$ of the M2 experiment become two times larger compared to the case of M1. $V_{6\to3}(t)$ for $x$=11.6 % (d) and $x$ = 21.5 % (f) are presented, which would possess larger signals on overall than $V_{4\to3}(t)$ when no circulating currents exist. In fact, the peak value of $V_{6\to3}(t)$ is only one sixth and one eighth of $V_{4\to3}(t)$ for $x$ = 11.6 % and $x$ = 21.5 %, respectively.

Fig. 5 (a) The M1 and M2 experiments for an n-doped GaAs single crystal. The $V_{4\to3}(t)$ [(a)] and $V_{6\to3}(t)$ [(b)] of the M1 experiment. Both signals are negative with fast and slow decay times. The magnitudes of $V_{4\to3}(t)$ and $V_{6\to3}(t)$ barely depend on the applied magnetic fields. (c) $V_{4\to3}(t)$ of the M2 experiment. Compared to the $V_{4\to3}(t)$ of the M1 experiment, this signal is reduced, which is a characteristic feature of the band "insulator".

Fig. 6 The experimental $V_{4\to3}(t)$ for $Bi_{1-x}Sb_x$ with $x$=11.6% and $x$ = 21.5% is normalized separately into a single universal curve. (a) Red line is a theoretical curve $A^* \mathrm{sec}\, h^2\left(\frac{t-t_0}{\tau}\right)$, fitted to the normalized $V_{4\to3}(t)$ of $Bi_{1-x}Sb_x$ with $x$=11.6 %. (b) The normalized $V_{4\to3}(t)$ of $Bi_{1-x}Sb_x$ with $x$ = 21.5 % and the theoretical fitting to $A^* \mathrm{sec}\, h^2\left(\frac{t-t_0}{\tau}\right)$ are shown. In contrast to $Bi_{1-x}Sb_x$ with $x$ = 11.6 %, the theoretical curve can fit the experimental data only slightly above the peak, beyond which the experimental data deviate from the theoretical curve. (c) The $V_{4\to3}(t)$ of $x$ = 21.5 % above the peak is fitted to $V_{4\to3}(t) = V_s \exp(-t/\tau_s) + V_l \exp(-t/\tau_l) + V_0$, extracting out the time constant of the slow relaxation. (d) The scaled $V_{4\to3}(t)$ data for Bi are shown, along with the theoretical curve $A^* \mathrm{sec}\, h^2\left(\frac{t-t_0}{\tau}\right)$ (the red line). The inset shows the $V_{4\to3}(t)$ data in a wider time window. Clearly, oscillating behaviours appear at $B$ = 0.8 T and $B$ = 4.0 T, where their oscillating amplitudes depend on applied magnetic fields.

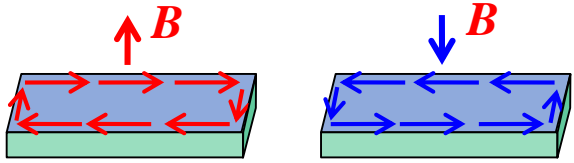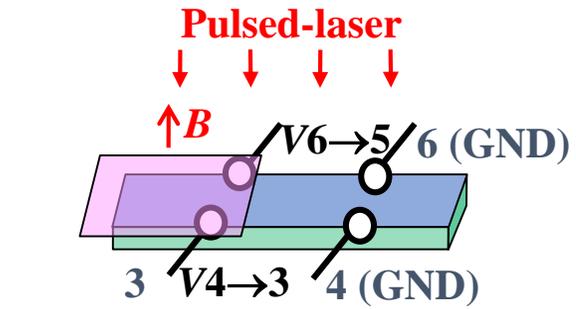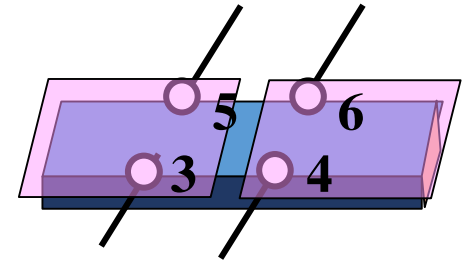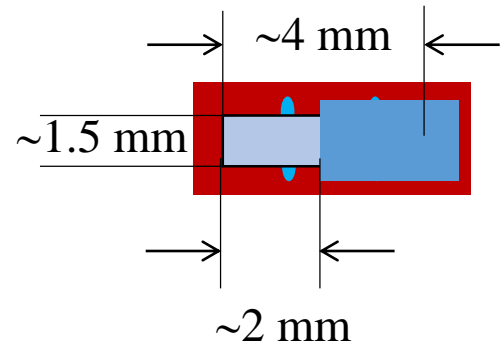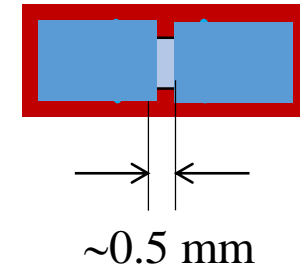

Fig. 1

**Fig. 2**

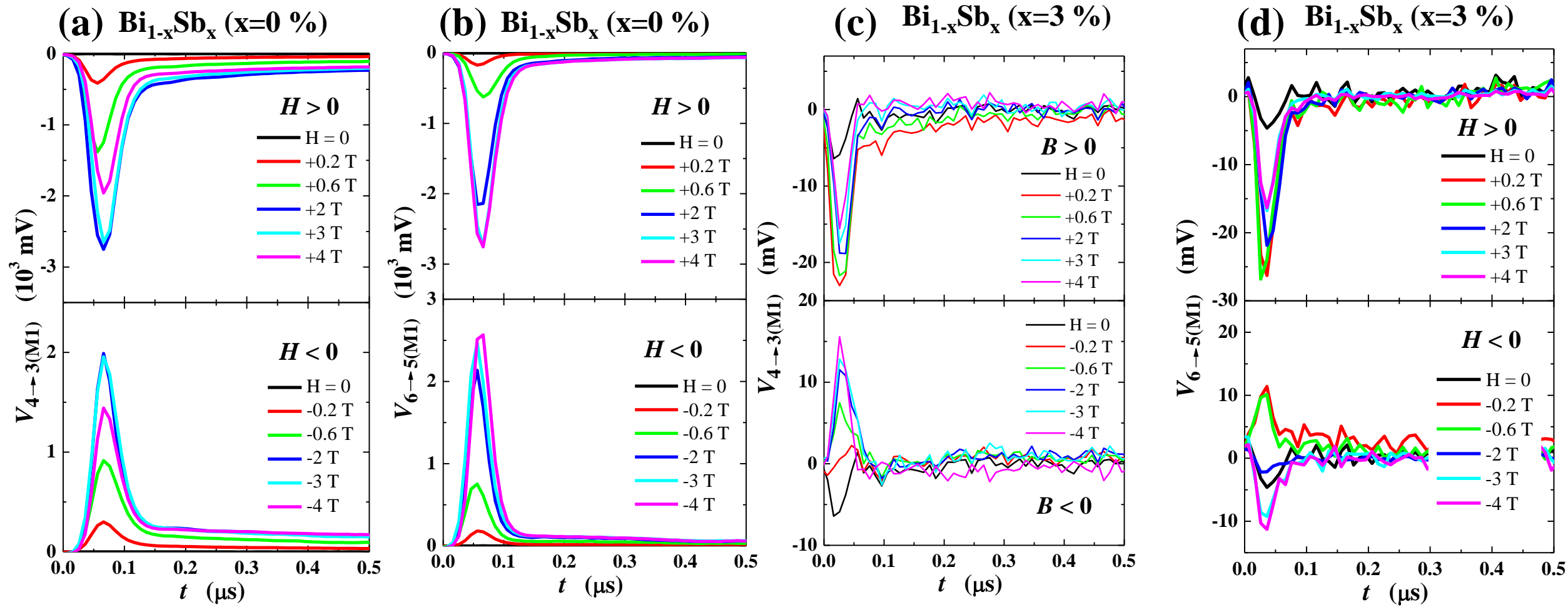

# Fig. 3

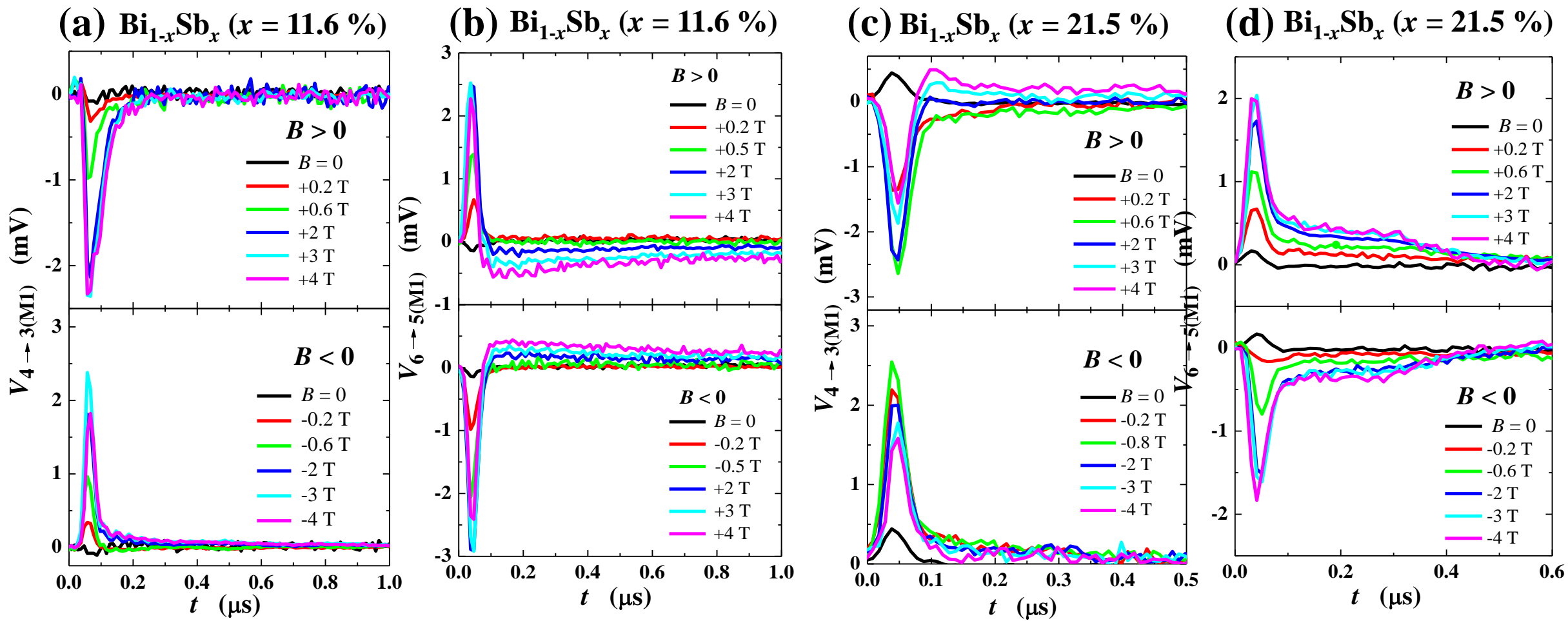

**Fig. 4**

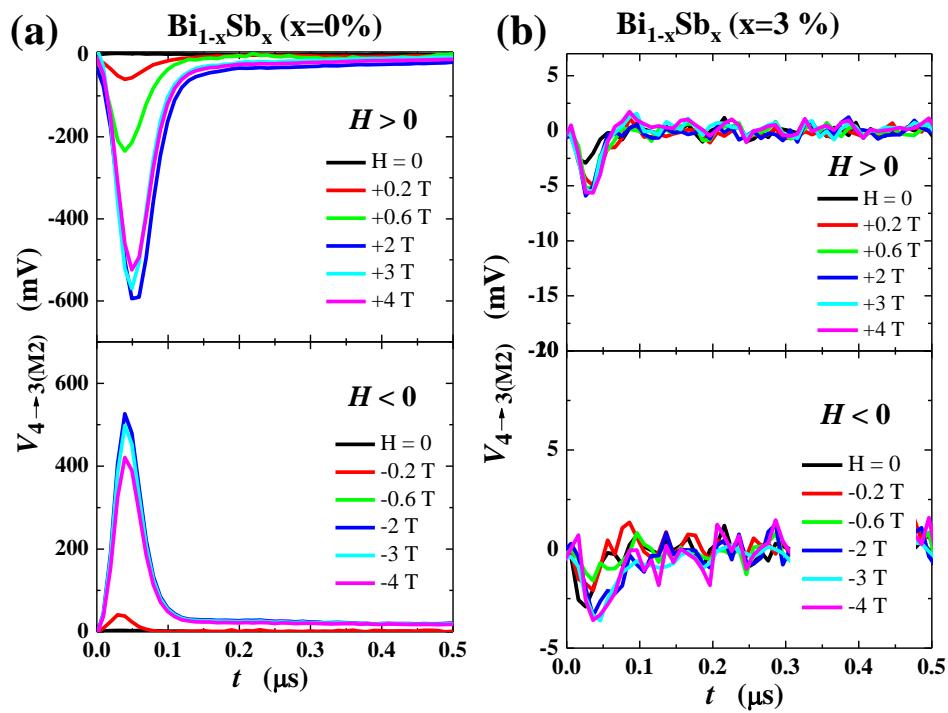
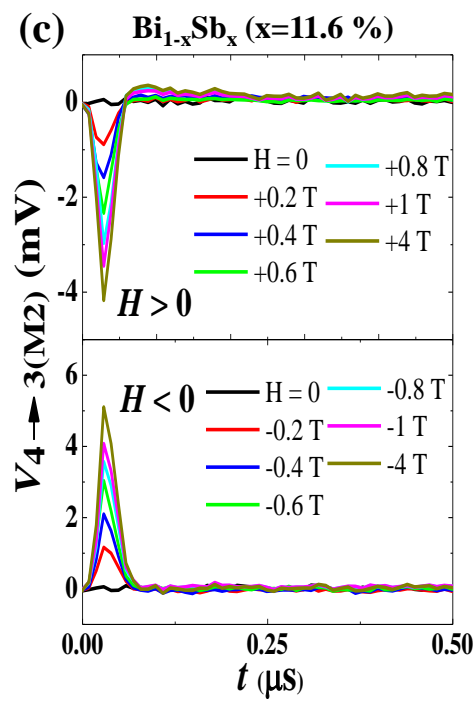
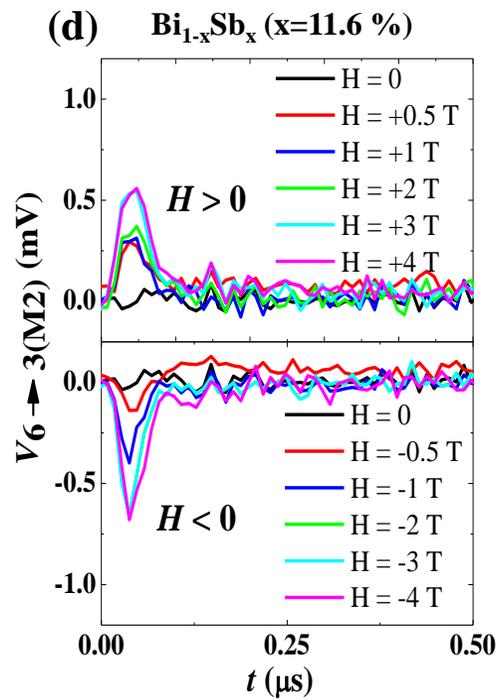
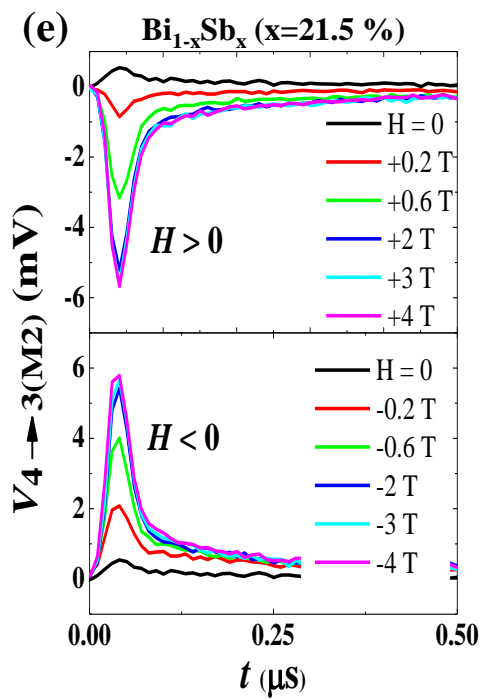
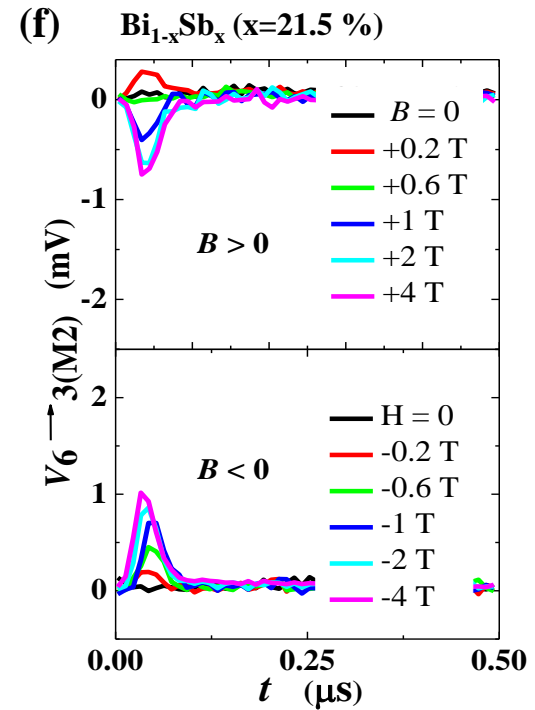

**Fig. 5**

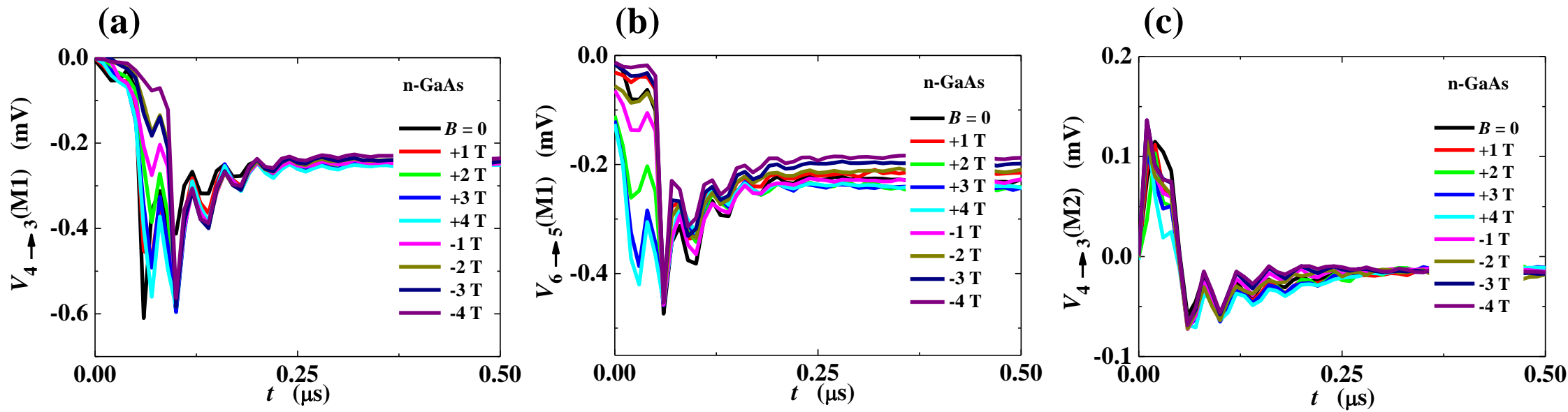

**Fig. 6**

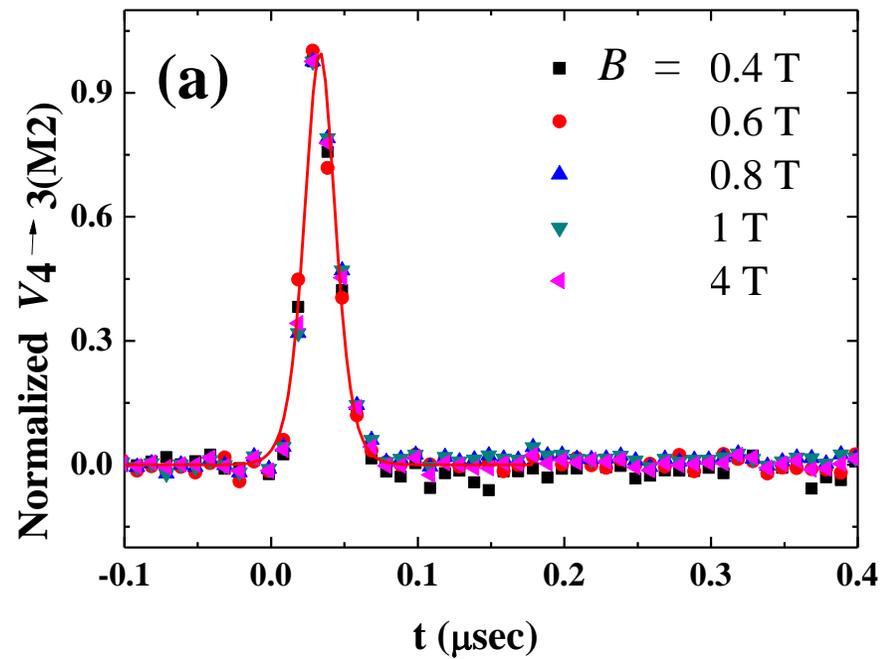
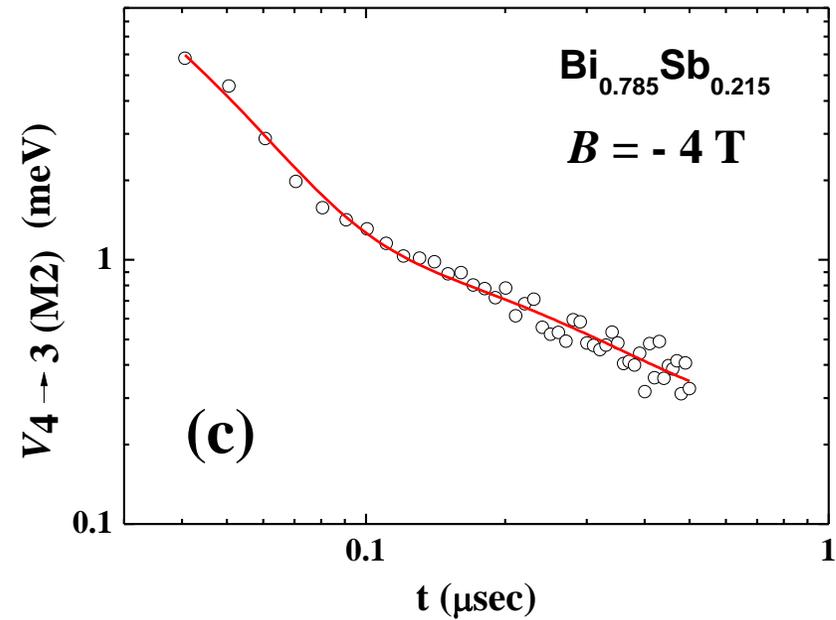
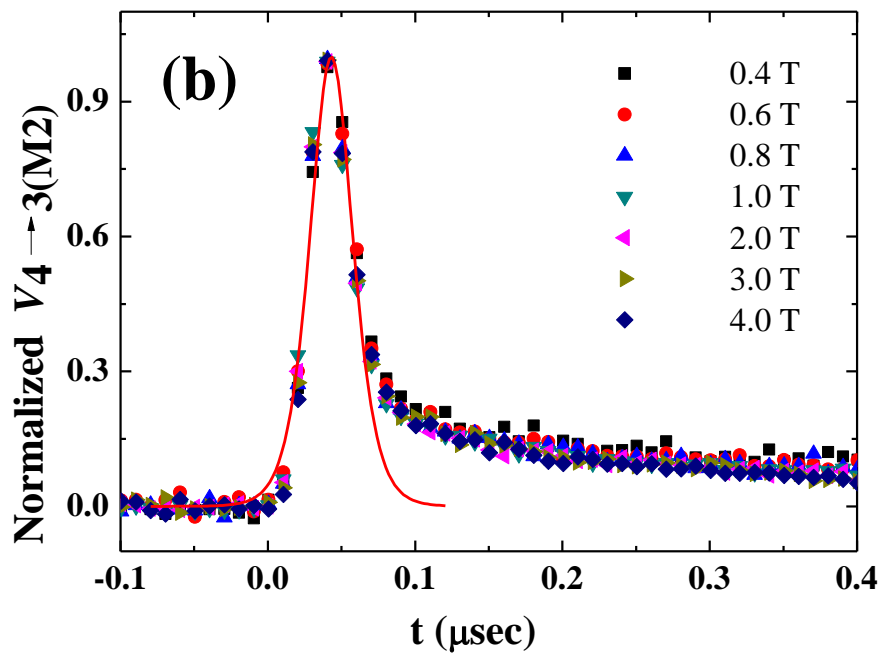
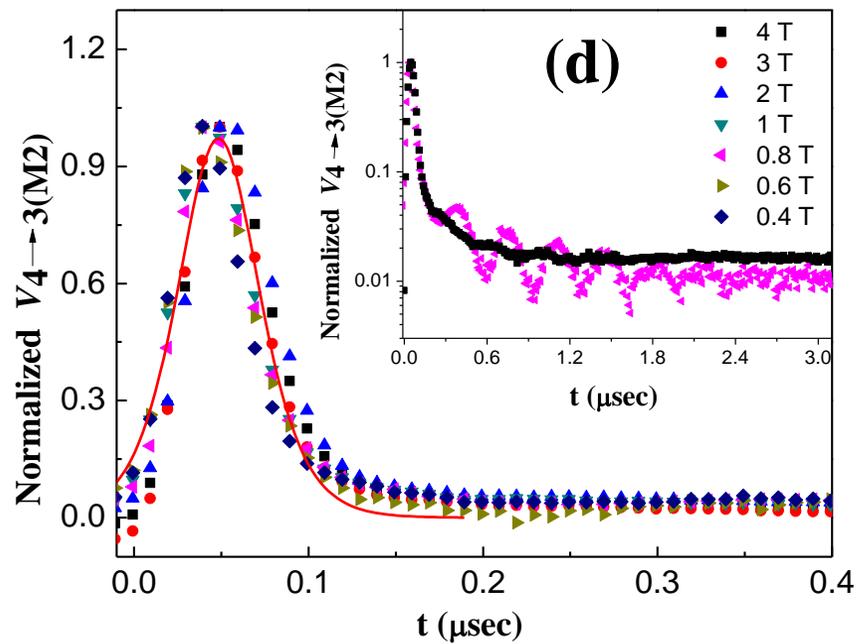